\documentclass{elsart}

\usepackage{graphicx}
\usepackage{pstricks}

\usepackage{dcolumn}
\newcolumntype{d}[1]{D{.}{.}{#1}}

\newcommand{\vG}{\mathbf{G}} 
\newcommand{\vk}{\mathbf{k}} 
\newcommand{\vq}{\mathbf{q}} 
\newcommand{\vr}{\mathbf{r}} 
\newcommand{\trans}{^{\rm T}} 
\newcommand{\sumlm}{\sum_{l=0}^{\infty}\sum_{m=-l}^l} 
\newcommand{\kLk}[1]{#1 \vk\trans \mathbf L #1 \vk} 
\newcommand{\Wlr}{W^{\rm lr}} 
\newcommand{\Wsr}{W^{\rm sr}} 
\newcommand{\oforder}{\mathrm O}
\newcommand{\Gk}{\vk+\vG}
\newcommand{\Gkp}{\vk+\vG'}

\begin{document}

\begin{frontmatter}
\title{Dielectric anisotropy in the \textit{GW} space-time method}
\author[FHI]{Christoph Freysoldt\thanksref{freyso}},
\author[FHI]{Philipp Eggert},
\author[FHI]{Patrick Rinke},
\author[FHI,Juelich]{Arno Schindlmayr},
\author[York]{R.W. Godby}, and
\author[FHI]{Matthias Scheffler}
\address[FHI]{Fritz-Haber-Institut der Max-Planck-Gesellschaft, Faradayweg 4--6, 14195 Berlin, Germany}
\address[Juelich]{Institut f\"ur Festk\"orperforschung, Forschungszentrum J\"ulich, 52425 J\"ulich, Germany}
\address[York]{Department of Physics, University of York, Heslington, York YO10 5DD, United Kingdom}

\thanks[freyso]{email: freyso@fhi-berlin.mpg.de}

\begin{abstract}
Excited-state calculations, notably for quasiparticle band structures,
are nowadays routinely performed within the $GW$ approximation for
the electronic self-energy. Nevertheless, certain numerical
approximations and simplifications 
are still employed in practice to make the computations feasible. 
An important aspect for periodic systems is 
the proper treatment of the singularity of the screened Coulomb 
interaction
in reciprocal space, which results from the slow $1/r$ decay in real space.
This must be done without introducing
artificial interactions between the quasiparticles and
their periodic images in repeated cells, which occur when
integrals of the screened Coulomb interaction are discretised in 
reciprocal space.
An adequate treatment of both aspects is crucial for 
a numerically stable computation of the self-energy.
In this article we build on existing schemes for isotropic screening and
present an extension for anisotropic systems. We also show
how the contributions to the dielectric function arising from the non-local 
part of the pseudopotentials can be computed efficiently.
These improvements are crucial for obtaining a fast 
convergence with respect to the number of points used for the Brillouin 
zone integration and prove to be essential to make $GW$ calculations 
for strongly anisotropic systems, such as slabs or multilayers, efficient.
\end{abstract}

\end{frontmatter}

\section{Introduction}

For describing quasiparticle excitations, as measured in direct and inverse
photoemission, many-body perturbation theory in the
 $GW$ approximation \cite{Hedin} has developed into the method of 
choice for weakly correlated solids and their surfaces.
In particular, the $GW$ approximation describes the quasiparticle band 
structures and band gaps for a large variety of semiconductors in good 
agreement with experimental results \cite{Aulbur,Rinke/etal:2005}.
For more correlated systems, it becomes necessary to go beyond $GW$,
but these schemes often include the $GW$ self-energy diagrams as lowest
order \cite{Biermann,Sun,Zein}.
Similarly, the Bethe--Salpeter approach to electron-hole excitations, as 
probed in optical absorption or electron energy-loss spectroscopy, builds 
on the $GW$ self-energy \cite{Onida}.

In the $GW$ approximation the frequency-dependent, non-local 
self-energy $\Sigma$ that connects
the independent-particle Green function $G_0$ with the interacting
one $G$ is given by $\Sigma = \mathrm iGW$, where $W$ is the 
dynamically screened Coulomb interaction.
The independent-particle starting point is typically chosen to be the 
Green function of a Kohn--Sham density-functional theory (DFT) calculation. 
Despite significant methodological progress
\cite{gwstPRL,Rohlfing,Aryasetiawan,Lebegue,Patterson,Tiago,Tiago2}
$GW$ calculations are still computationally demanding, and
their application is limited
to relatively small system sizes. So far all implementations employ 
a number of additional
simplifications to reduce the computational cost. Some of these are
motivated by physical considerations, such as plasmon-pole models 
\cite{HybertsenLouie} or
model dielectric functions
\cite{LevineLouie,modelDielec3,modelDielec,modelDielec2},
while others appear as purely
mathematical ``tricks'' to improve the numerical stability or efficiency.
Often, the validity and usefulness of 
a specific approach depends on the physical system under consideration.
An important aspect in every $GW$ implementation that uses 
reciprocal space is the treatment of the singularity at 
$\vk\rightarrow\bf 0$ in
the bare and screened interaction for non-metallic systems.
This singularity is integrable, and many different schemes have 
been developed for these $\vk$-space integrals 
\cite{Lebegue,HybertsenLouie,Rieger,Hott,Pulci,Wenzien}.
The general idea is to describe the singular part by a model function that 
can be handled analytically, so that the remainder is sufficiently smooth
for a numerical treatment.
In physical terms, the $\vk\rightarrow \bf 0$ behaviour determines the 
long-range part of the interaction. The discretisation of the 
reciprocal-space integrals can then be interpreted as introducing an artificial 
supercell periodicity in real space. The periodic images of the 
quasiparticles give rise to an infinite self-interaction due to the $1/|\vr|$
tail of the interaction.
Any scheme that integrates the singularity analytically is therefore strictly 
equivalent to modifying the long-range behaviour of the interaction
such that a quasiparticle does not interact with its periodic images or
that the interaction decays faster than $1/|\vr|^2$.
In other words, different integration schemes correspond to particular 
modifications of the long-range tail and vice versa.

In most schemes screening is assumed to be isotropic
at the length scale given by the inverse of the smallest 
non-zero $\vk$-vector, which is appropiate for most bulk materials
but may fail in systems with an appreciable anisotropy, such as
superlattices or layered materials, as well as 
in supercell approaches for low-dimensional materials
like clusters, molecules, nanowires, films, or surfaces.
An obvious way to avoid these spurious interactions for systems with broken
translational symmetry is to abandon the concept of 
periodic boundary conditions altogether in the relevant directions
and perform the calculation entirely in real space.
For semi-infinite jellium surfaces such a $GW$ embedding scheme has been
successfully implemented \cite{Fratesi1,Fratesi2}. 
Its extension to realistic
surfaces, however, is computationally still too expensive. 
A real-space implementation for finite systems has also been
reported \cite{Tiago,Tiago2}, but its applicability to systems with periodicity
in one or more directions remains to be shown.

Staying with the repeated-cell approach,
we will show in the following how it is possible to incorporate
the anisotropy in the treatment of the singularity
in the $GW$ space-time method \cite{Rieger}.
In addition to the equations that we have implemented
we derive exact expressions that allow us to 
discuss other algorithms in comparison. Furthermore, we will show
that the proposed modifications considerably improve the
convergence behaviour with respect to the number of 
$\vk$-points, which is the natural parameter
associated with the singularity treatment.
In practice the $GW$ approximation is often applied 
non-self-consistently by constructing the screened interaction
as well as the self-energy from the independent-particle Green 
function $G_0$. However, the behaviour of the 
anisotropy discussed in this article applies equally to the fully
self-consistent $GW$ approach. 
For simplicty we will therefore focus on the
non-self-consistent case and indicate differences whenever they apply.
In the interest of readability we will also refrain from introducing 
different symbols to distinguish between the self and non-self-consistent case.

This paper is organised as follows. In Section~\ref{sec:gwst_outline} 
we briefly describe the computation of the
self-energy in the space-time method. In 
Section~\ref{sec:anisotropy} we
explain how the anisotropy is accounted for;
the detailed derivation of the well-known anisotropic
equations is presented in Appendix \ref{ap:dielec} for completeness.
In Section~\ref{sec:results} we demonstrate the
improved $\vk$-point convergence behaviour resulting from our modifications
before summarising our results in Section~\ref{sec:conclusions}.
In Appendix~\ref{app:angularExpansion} we have collected the 
spherical-harmonics expansion of several vector quantities that 
appear in our derivations. Finally,
an efficient implementation of the contribution from 
Kleinman--Bylander-type non-local pseudopotentials \cite{KleinmanBylander},
which enter the expressions for
the anisotropy, is presented in Appendix~\ref{app:nonlocal_momentum}.
Unless otherwise indicated, we use Hartree atomic units.

\section{Outline of the \textit{GW} space-time method}
\label{sec:gwst_outline}

The $GW$ space-time method has been presented in detail elsewhere
\cite{Rieger,gwstGL}.
We will therefore only summarise the steps to construct the self-energy
from the output of a preceding DFT calculation. Assuming
a non-magnetic systems for simplicity (the extension to a spin-dependent 
Green function is straight-forward), the computational steps are:
\begin{enumerate}
\item
Construction of the non-interacting Green function $G$ in real space and 
imaginary time from the Kohn--Sham eigenfunctions $\varphi_{n\vk}$ and 
eigenvalues $\epsilon_{n\vk}$ (the Fermi level is set as the energy zero)
\begin{equation}
G(\vr,\vr';{\rm i}\tau) = 
\mathrm i \frac{\Omega}{(2\pi)^3}\int_{\rm BZ}\!\mathrm d^3 k\left\{\begin{array}{ll}
\sum\limits_n^{\rm occ} \varphi_{n\vk}(\vr) \varphi_{n\vk}^*(\vr')
\mathrm e^{- \epsilon_{n{\vk}}\tau}, & \tau < 0,\\
-\!\!\sum\limits_n^{\rm unocc} \varphi_{n\vk}(\vr) \varphi_{n\vk}^*(\vr')
\mathrm e^{-\epsilon_{n{\vk}}\tau}, & \tau > 0,\\
\end{array}\right.
\end{equation}
where $\Omega$ denotes the unit-cell volume and the integral over
$\bf k$ runs over the first Brillouin zone,
\item
\label{step:pol}
formation of the irreducible polarisability $P$ in the 
random-phase approximation in real space and
imaginary time
\begin{equation}
P(\vr,\vr';{\rm i}\tau) = -2\mathrm i 
G(\vr,\vr';{\rm i}\tau) G(\vr',\vr;-{\rm i}\tau)
\; ,
\end{equation}
\item Fourier transformation of $P$ to reciprocal space
\begin{equation}
P_{\vG\vG'}(\vk,{\rm i}\tau) = \frac{1}{\Omega}
\int\!\mathrm d^3r\int\!\mathrm d^3r'~ P(\vr,\vr';{\rm i}\tau)
 \mathrm e^{-{\rm i}(\Gk) \cdot\vr + {\rm i}(\Gkp)\cdot\vr'} 
\end{equation}
and to imaginary frequency,
\item construction of the symmetrised dielectric matrix
in reciprocal space
\begin{equation}
\tilde \varepsilon_{\vG\vG'}(\vk, {\rm i}\omega) = \delta_{\vG\vG'} - 
\frac{4\pi}{|\Gk||\Gkp|} P_{\vG\vG'} (\vk, {\rm i}\omega)
\; ,
\label{eq:eps_form}
\end{equation}
\label{step:eps_form}
\item
inversion of the symmetrised dielectric matrix for each $\vk$-point
and each imaginary frequency,
\label{step:eps_inv}
\item calculation of the screened Coulomb interaction in reciprocal
space
\begin{equation}
W_{\vG\vG'}(\vk, {\rm i}\omega) = 
\frac{4\pi}{|\Gk||\Gkp|} \tilde\varepsilon^{-1}_{\vG\vG'} (\vk, {\rm i}\omega)
\; ,
\label{eq:define_W}
\end{equation}
\label{step:W_form}
\item Fourier transformation of $W$ to imaginary time and to real space
\label{step:W_fourier}
\begin{equation}
W(\vr,\vr';{\rm i}\tau) = \frac{1}{(2\pi)^3}\int_{\rm BZ}\!\!\mathrm d^3k \sum_{\vG,\vG'}
W_{\vG\vG'}(\vk,{\rm i}\tau) 
\mathrm e^{{\rm i}(\Gk) \cdot\vr - {\rm i}(\Gkp)\cdot\vr'} 
\label{eq:define_WinR}
\; ,
\end{equation}
\item computation of the self-energy in real space and imaginary time
\label{step:last}
\begin{equation}
\Sigma(\vr, \vr'; {\rm i}\tau) = \mathrm i G(\vr, \vr'; {\rm i}\tau) W(\vr, \vr'; {\rm i}\tau)
\; .
\end{equation}
\end{enumerate}
The Coulomb singularity appears explicitly for $\vG=\bf 0$ or $\vG'= \bf 0$
as $\vk\rightarrow \bf 0$ in steps~\ref{step:eps_form} and~\ref{step:W_form};
in the actual implementation, however, it is treated 
in steps~\ref{step:eps_inv} and~\ref{step:W_fourier}
for numerical reasons.
The anisotropy enters the scheme naturally through the construction of
the dielectric matrix and must be taken
fully into account in the screened interaction.

The quasiparticle energies are obtained by computing the matrix elements 
of the self-energy
$\langle\varphi_{n\vk}|\Sigma({\rm i}\tau)|\varphi_{n\vk}\rangle$ 
on the imaginary time axis, which are then Fourier-transformed to imaginary
frequency and analytically continued to the real frequency axis.
Approximating the quasiparticle by the DFT Kohn-Sham wavefunctions 
finally gives the quasiparticle energies $\epsilon^{\rm qp}_{n\vk}$ as
solutions of the quasiparticle equation
\begin{equation}
\label{eq:qpe}
\epsilon^{\rm qp}_{n\vk} = \epsilon_{n\vk} + 
\langle\varphi_{n\vk}|
\Sigma(\epsilon^{\rm qp}_{n\vk})-V_{\rm xc}
|\varphi_{n\vk}\rangle
\; ,
\end{equation}
where $V_{\rm xc}$ is the exchange-correlation potential used in 
the underlying DFT calculation.
The details of the analytic continuation and the solution of 
Equation (\ref{eq:qpe}) have been described elsewhere \cite{Rieger}.
Self-consistency in $GW$ would be achieved by entering step 
\ref{step:pol} with a new Green
function obtained from solving Dyson's equation
$G=G_0+G_0\Sigma G$ after step 8 and iterating steps 
\ref{step:pol}--\ref{step:last}. 

\section{Treatment of the anisotropy}
\label{sec:anisotropy}

\subsection{Anisotropy in the screened interaction}
As shown in Appendix \ref{ap:dielec} the
head ($\vG=\vG'={\bf 0}$) and the wings ($\vG={\bf 0}$ or $\vG'={\bf 0}$) of the 
dielectric matrix close to the $\Gamma$-point, i.e.,  for
$\vk \rightarrow{\bf 0}$, depend on the direction in which this limit 
is taken. We denote this dependence by the spatial 
angle $\Omega_\vk$, and the corresponding normalised direction vector by $\hat \vk$.
For simplicity, the imaginary frequency argument ${\rm i}\omega$
is omitted in the following.
The directional dependence at the $\Gamma$-point is present in the 
whole inverse dielectric matrix, i.e., head, wings, and body.
By block-wise inversion \cite{PickCohenMartin} it is easily shown that the
head of the inverse symmetrised dielectric matrix takes the form
(cf. Appendix \ref{ap:dielec})
\begin{equation}
\label{eq:head_aniso}
\tilde \varepsilon^{-1}_{\bf 00}(\Omega_\vk)  =  \frac{1}{\kLk{\hat}}
\; ,
\end{equation}
where the matrix $\mathbf L$ is the macroscopic dielectric tensor.
We note that in most other implementations, in which this anisotropy 
has been considered, 
such as \cite{Pulci,Wenzien}, but not \cite{Hott},
the right-hand side of Equation (\ref{eq:head_aniso}) has been replaced by 
the expression $\hat \vk\trans \mathbf L^{-1} \hat \vk$ without formal
justification.

Correspondingly, the wings can be expressed as
\begin{equation}
\tilde \varepsilon^{-1}_{\vG \bf 0}(\Omega_\vk)=
- \tilde \varepsilon^{-1}_{\bf 00}(\Omega_\vk)  
\Big[\hat \vk \cdot \mathbf S(\vG)\Big]
\; .
\end{equation}
The vector $\mathbf S(\mathbf G)$ is defined in Equation (\ref{eq:definitionS}).
For the remainder of this article we will restrict the discussion of the wings to 
the expression  for $\vG'={\bf 0}$ since the case $\vG={\bf 0}$ is trivially obtained 
from the symmetry relation
\begin{equation}
\tilde\varepsilon^{-1}_{\bf 0 G'}(\Omega_\vk) = 
[\tilde\varepsilon^{-1}_{\bf G' 0}(\Omega_\vk)]^*
\; .
\end{equation}
Finally, the body is given by
\begin{equation}
\tilde \varepsilon^{-1}_{\bf GG'}(\Omega_\vk) =
B^{-1}_{\vG\vG'}
+ \tilde \varepsilon^{-1}_{\bf 00}(\Omega_\vk)  
\Big[ \hat \vk \cdot \mathbf S(\vG) \Big]
\Big[ \hat \vk \cdot \mathbf S(\vG') \Big]^*
\; ,
\label{eq:epsi_body}
\end{equation}
where $\mathbf B$ denotes the body of the symmetrised dielectric 
matrix as defined in Equation (\ref{eq:defineHWB}).
Expressions (\ref{eq:head_aniso})-(\ref{eq:epsi_body}) imply that
each element of the inverse dielectric matrix is, in general,
not analytic at $\vk={\bf 0}$, i.e., it does not have a unique limit for 
$\vk \rightarrow {\bf 0}$. This was already recognised more than 30 years ago
by Pick et al. \cite{PickCohenMartin}, but the treatment of
this non-analytic behaviour in $GW$ implementations
has not been discussed widely in the literature.
Hybertsen and Louie address the problem in the appendix of 
\cite{HybertsenLouie}, but in the actual calculation they neglect
the non-analytic part, arguing that the error can be made negligibly
small with a sufficiently high number of $\vk$-points.
So far the anisotropy has only been considered 
for the head element in connection with the treatment
of the Coulomb singularity \cite{Hott,Pulci,Wenzien}. 
We will return to this point later in Section~\ref{sec:reciprocalCodes}.

Combining Equations (\ref{eq:head_aniso}) to (\ref{eq:epsi_body})
with Equation (\ref{eq:define_W}), we obtain the screened 
interaction for $\vk\rightarrow\bf0$
\begin{eqnarray}
W_{\bf 00}(\vk)  &\rightarrow&  
\frac{4\pi}{|\vk|^2}\tilde\varepsilon^{-1}_{\bf 00}(\Omega_{\vk})
\label{eq:Whead}
\; ,
\\
W_{\vG \bf 0}(\vk)&\rightarrow&
- \frac{4\pi}{|\bf k||G|}  \tilde \varepsilon^{-1}_{\bf 00}(\Omega_\vk)  
\Big[\hat \vk \cdot \mathbf S(\vG)\Big]
\; ,
\\
W_{\bf GG'}(\vk) &\rightarrow&
\frac{4\pi}{|\bf G||G'|}\left(
B^{-1}_{\vG\vG'}
+ \tilde \varepsilon^{-1}_{\bf 00}(\Omega_\vk)  
\Big[ \hat \vk \cdot \mathbf S(\vG) \Big]
\;
\Big[ \hat \vk \cdot \mathbf S(\vG') \Big]^*
\right)
\; .
\end{eqnarray}

The presence of the singularity at $\bf G=G'=0$ 
(Equation~(\ref{eq:Whead})) necessitates a special numerical treatment.
In the space-time method,
this problem is solved by splitting off a long-range part $\Wlr$
with the appropiate $\vk \rightarrow \mathbf 0$ behaviour, which is chosen
such that its Fourier transform can be computed semi-analytically
as described in detail in the following section.
The remaining short-range part $\Wsr = W - \Wlr$ can then safely be 
treated numerically since it is no longer singular.

The next step in the space-time method is the Fourier transformation of $W$ to
real space (Equation (\ref{eq:define_WinR}));
in reciprocal-space algorithms it is the construction of
the matrix elements of $\Sigma$. Both approaches involve an integration over 
the Brillouin zone after multiplication by an analytic function 
$a_{\bf GG'}(\vk)$, and it is in this integration that the anisotropy
must be taken into account.
In practice, these integrals are usually discretised,
which we express formally by partitioning the Brillouin zone
into subzones $\mathrm Z_i$ with volume $V_i$:
\begin{equation}
\int_{\rm BZ}\!\!\mathrm d^3 k~ W_{\bf GG'}(\vk) a_{\bf GG'}(\vk)
= \sum_i \int_{\mathrm Z_i}\!\mathrm d^3 k~ W_{\bf GG'}(\vk) a_{\bf GG'}(\vk)
\label{eq:brilldiscret}
\; .
\end{equation}
We denote the subzone that contains the $\Gamma$-point
by $\mathrm Z_\Gamma$ and assume that it
has inversion symmetry about $\vk={\bf 0}$.
While the subzone integrals for $i\ne\Gamma$ can be
approximated by
\begin{equation}
\int_{\mathrm Z_i}\!\mathrm d^3 k~ W_{\bf GG'}(\vk) a_{\bf GG'}(\vk) \approx 
V_i  W_{\bf GG'}(\vk_i) a_{\bf GG'}(\vk_i)
\label{eq:intz_ngamma}
\; ,
\end{equation}
where $\vk_i$ is a representative point for the subzone $\mathrm Z_i$ --
usually its centre -- the integrals over $\mathrm Z_{\Gamma}$
require a special treatment due to the non-analyticity of $W$ at $\Gamma$.

In principle, even the singularity for $\bf G=G'=0$ can be treated
in this way. Since it is illuminating to discuss existing isotropic and 
anisotropic singularity integration schemes in terms of approximations to an 
exact expression, we present the corresponding equations for 
reciprocal space algorithms in Section \ref{sec:reciprocalCodes}.
In the space-time method, on the other hand, the separation 
into $\Wlr$ and $\Wsr$ is more efficient than the direct approach, 
because the analytic function 
$a_{\bf GG'}(\vk)$ for the Fourier transformation 
\begin{equation}
a_{\bf GG'}(\vk) 
= \mathrm e^{\mathrm i(\Gk)\cdot\vr - \mathrm i (\Gkp)\cdot\vr'}
\label{eq:aGGk_FT}
\end{equation}
depends on $\vr$ and $\vr'$ and would thus require the computation of
the integrals over $\mathrm Z_\Gamma$ in Equation (\ref{eq:brilldiscret})
for every $\vr$ and $\vr'$. 

Since the
long-range part of the interaction yields a significant contribution
to the quasiparticle energies, an accurate treatment of its anisotropy
is very important and is therefore described first in 
Section \ref{sec:head}.
In Section \ref{sec:wings} we show that the apparent 
$1/|\vk|$ singularity of the wings does not cause numerical problems
for the integrals over $Z_\Gamma$, and in Section \ref{sec:body} the
computation of the integrals for the body is presented.
 
\subsection{Treatment of the head}
\label{sec:head}

In the space-time method
the head of the inverse dielectric matrix is used to define the
long-range part of the screened interaction. 
For this purpose, we extend Equation~(\ref{eq:Whead}) to 
$\vG=\vG'\ne\mathbf 0$ and define the long-range part for
all $\vk$ in the Brillouin zone as
\begin{equation}
\Wlr_{\bf GG'}(\vk) = \frac{4\pi}{(\Gk)\trans  \mathbf L (\Gk)} \delta_{\vG\vG'}
\; .
\end{equation}
For numerical reasons we subtract the long-range part at the level
of the inverse dielectric matrix after applying the body corrections described
in Section \ref{sec:body} for $\vk=\mathbf 0$, and 
compute $\Wsr$ from this modified entity according to
\begin{eqnarray}
\tilde\varepsilon_{\bf GG'}^{-1,{\rm sr}}(\vk) &:=& 
\tilde\varepsilon_{\bf GG'}^{-1}(\vk)
- \frac{|\Gk|^2}{({\Gk})\trans \mathbf L ({\Gk})}\delta_{\bf GG'}
\; ,
\\
\Wsr_{\bf GG'}(\vk) &= & \frac{4\pi}{|\Gk||\Gkp|}
\tilde\varepsilon_{\bf GG'}^{-1,{\rm sr}}(\vk)
\; .
\end{eqnarray}
By expanding the angular dependence of $\Wlr$ into spherical harmonics (cf. Appendix~\ref{app:angularExpansion})
\begin{equation}
\Wlr_{\bf GG'}(\vk) = \sumlm H_{lm}
 \frac{4\pi}{|\Gk|^2}  \delta_{\bf GG'}Y_{lm}(\Omega_{\Gk})
\label{eq:W_hlm}
\; ,
\end{equation}
the Fourier transformation of $\Wlr$ can be performed analytically. 
Only even $l$ contribute to
the sum because the coefficients $H_{lm}$ vanish for
odd $l$.
Making use of the expansion of a plane wave \cite{Jackson} 
in spherical harmonics $Y_{lm}$ and spherical Bessel functions $j_l$,
\begin{equation}
\mathrm e^{\mathrm i\vk\cdot\vr}
=4\pi\sumlm \mathrm i^l j_l(kr)
    Y_{lm}(\Omega_{\vr}) Y_{lm}^{*}(\Omega_{\vk})
\end{equation}
we arrive at
\begin{equation}
\Wlr(\vr,\vr') = 
\sumlm
c_l \mathrm i^l H_{lm} Y_{lm}(\Omega_{\vr-\vr'}) \frac{1}{|\vr-\vr'|}
\label{eq:Wl_in_r}
\; .
\end{equation}
The coefficients $c_l$ for even $l$ are defined as
\begin{eqnarray}
 c_{l}    &=& \frac{2}{\pi}\int_{0}^{\infty}\!\!\mathrm dx~j_{l}(x)
   =\frac{(l-1)!!}{l!!}
\end{eqnarray}
with $n!! = n (n-2)(n-4) \cdots$.
In practice we truncate the sum in Equation~(\ref{eq:Wl_in_r}) at
finite $l=l_{\rm max}$ (cf. Section \ref{sec:conv_L}).

For numerical convenience $\Sigma$ is split into a static exchange part
$\Sigma_{\rm x}=i G v$ and a frequency-dependent correlation part 
$\Sigma_{\rm c} = i G (W - v)$ in the space-time method \cite{Rieger}.
This is achieved by subtracting the unscreened Coulomb interaction $v$
from $\Wlr$ in its angular expansion (\ref{eq:W_hlm}),
i.e., $1/\sqrt{4\pi}$ is subtracted from $H_{00}$ for each imaginary frequency.
Furthermore, the transformation from imaginary frequency to imaginary time 
is performed on the expansion coefficients 
$H_{lm}({\rm i}\omega)$ directly,
and we obtain $(\Wlr - v)$ according to Equation~(\ref{eq:Wl_in_r})
with the expansion coefficients in imaginary time $H_{lm}({\rm i}\tau)$.

A proper treatment of the anisotropy in the 
long-range part of the screened interaction is crucial to obtain 
converged results. This is easily illustrated in the space-time
method:
For non-local operators like $W$ or $\Sigma$ the density of the $\vk$-point 
sampling determines the range of the non-locality in real space. For example,
a 4$\times$4$\times$4 $\vk$-grid  corresponds to a maximum non-locality range
or \emph{interaction cell} of 4 real-space unit cells in each dimension.
If parts of the long-range interaction remain in $\Wsr$ for small 
but finite $\vk$, the tails of $\Wsr$ extend over the boundary of the 
interaction cell
and will be folded back in the numerical Fourier transformation
when applying the periodic boundary conditions. 
Since the size of the interaction cell 
is determined by the $\vk$-point sampling, 
an inadequate treatment of the long-range part would result
in an unsatisfactory $\vk$-convergence behaviour.

\subsection{Treatment of the wings}
\label{sec:wings}
The wings are antisymmetric with respect to $\vk$, i.e.,
\begin{equation}
W_{\bf G0}(\vk) = - W_{\bf G0}(-\vk)\;.
\end{equation}
Hence we can write the $\Gamma$-point contribution as
\begin{eqnarray}
&&\int_{\mathrm Z_\Gamma}\!\mathrm d^3 k~ 
W_{\bf G0}(\vk) a_{\bf G0}(\vk)
\nonumber\\
&=&\frac 12 \left(\int\nolimits_{\mathrm Z_\Gamma}\!\mathrm d^3 k~ 
W_{\bf G0}(\vk) a_{\bf G0}(\vk)
+ \int\nolimits_{\mathrm Z_\Gamma}\!\mathrm d^3 k~
W_{\bf G0}(-\vk) a_{\bf G0}(-\vk)
\right)
\nonumber\\&=&
- \frac{4\pi}{|\vG|}  
\int_{\mathrm Z_\Gamma}\!\mathrm d^3 k~ 
 \tilde \varepsilon^{-1}_{\bf 00}(\Omega_\vk)  
\Big[\hat \vk \cdot \mathbf S(\vG)\Big]
 \left[\hat \vk \cdot \nabla_{\vk} a_{\bf G0}(\Omega_\vk)\big|_{\vk = \mathbf 0} +
\oforder(|\vk|^2) \right]
\label{eq:wing_integrand}
\; ,
\end{eqnarray}
where we have made use of the Taylor expansion of the analytic 
function $a_{\bf G0}(\vk)$.
The important benefit of this reformulation is that no term in the
integrand is singular, and therefore we do not expect
any numerical difficulties close to the $\Gamma$-point.
This applies equally to the Fourier
transformation in the space-time method as well as to the construction
of the self-energy in reciprocal-space approaches.

In practice we neglect the wing contributions from the $\Gamma$-point 
in the Fourier transformation,
because evaluating Equation (\ref{eq:wing_integrand}) 
for each $\vr$ and $\vr'$ would be computationally very demanding.
The associated error scales as $V_{\Gamma}$. It is thus
automatically controlled by the standard $\vk$-point convergence
tests, since $V_{\Gamma}$ is inversely proportional to
the number of points in our regular $\vk$-point grid.
Test calculations 
indicate that the overall convergence behaviour shows no significant
improvement for a more sophisticated treatment of the wings.
We note that in contrast to head and body the isotropic
average for the wing contribution vanishes, i.e., 
no analytic contribution has to be considered when the non-analytic
part is neglected.

\subsection{Treatment of the body}
\label{sec:body}
For $|\vk| \ll |\vG|$ we see from Equation (\ref{eq:aGGk_FT})
that we can approximate $a_{\bf GG'}(\vk) \approx a_{\bf GG'}({\bf 0})$.
Similar considerations apply to the analytic function in 
reciprocal-space approaches.
We can then express
\begin{equation}
\int_{\mathrm Z_\Gamma} \!\mathrm d^3 k~ W_{\bf GG'}(\vk) a_{\bf GG'}(\vk)
\approx a_{\bf GG'}({\bf 0}) \int_{\mathrm Z_\Gamma} \!\mathrm d^3 k ~
\frac{4\pi \tilde \varepsilon^{-1}_{\bf GG'}(\Omega_\vk)}{|\vG||\vG'|}
\end{equation}
as a discretised contribution analogous to 
Equation~(\ref{eq:intz_ngamma}) with an averaged anisotropy
\begin{equation}
\tilde\varepsilon^{-1}_{\bf GG'}({\bf 0}) :=
\frac{1}{V_\Gamma}
\int_{\mathrm Z_\Gamma}\!\mathrm d^3 k ~
\tilde \varepsilon^{-1}_{\bf GG'}(\Omega_\vk) 
\; .
\end{equation}

We will now show how this average can be computed 
efficiently. To this end we rewrite the integral in polar coordinates
as
\begin{equation}
\tilde\varepsilon^{-1}_{\bf GG'}(\mathbf 0) = \int \!\mathrm d\Omega_\vk~
w(\Omega_\vk) 
 \tilde\varepsilon^{-1}_{\bf GG'}(\Omega_\vk)
\quad\textnormal{with}\quad
w(\Omega_\vk)= 
\frac{1}{V_\Gamma}\!\!\int_0^{k_{\rm max}(\Omega_\vk)} \!\!k^2\,\mathrm dk
\; ,
\label{eq:bodyaverage}
\end{equation}
where $k_{\rm max}(\Omega_\vk)$ is the distance from the centre to the
surface of $Z_\Gamma$ in the direction of $\Omega_\vk$, and
$w(\Omega_\vk)$ acts as an angular weight function 
that takes the shape of the $\Gamma$-zone element into account
and may be subject to additional approximations, e.g., for spherical 
averages it is simply a constant $w(\Omega_\vk)=1/4\pi$.

Inserting Equation (\ref{eq:epsi_body}), we obtain a very simple expression 
\begin{eqnarray}
\tilde \varepsilon^{-1}_{\bf GG'}(\mathbf 0) 
&=&
B^{-1}_{\vG\vG'} +
\sumlm \sum_{m'=-1}^1\sum_{m''=-1}^1 \overline H_{lm}  
S_{m'}(\vG)  S^*_{m''}(\vG')
\nonumber\\&&\times%
\int \! \mathrm d\Omega_\vk~ Y_{lm}(\Omega_\vk) 
Y_{1m'}(\Omega_\vk)  Y^*_{1m''}(\Omega_\vk)%
\label{eq:body_result}
\end{eqnarray}
when expanding 
$[w(\Omega_\vk) \tilde\varepsilon^{-1}_{\bf 00}(\Omega_\vk)]$
as well as $\hat\vk \cdot \mathbf S(\vG)$ in spherical harmonics 
as described in Appendix~\ref{app:angularExpansion}.
The angular integrals in Equation (\ref{eq:body_result}) are 
nothing but the
Clebsch--Gordan coefficients for the spherical harmonics $(lm\;1m'|1m'')$.
From the properties of the Clebsch--Gordan coefficients 
\cite{Abramowitz} it follows that
only a small number of non-zero terms contribute to 
Equation~(\ref{eq:body_result}),
namely $l=0$ with 3 terms and $l=2$ with 9 terms. We refer to these
12 terms as ``body corrections''.
In the original space-time implementation \cite{Rieger},
Equation~(\ref{eq:epsi_body}) was averaged 
over the Cartesian directions, which is equivalent to including 
only the $l=0$ terms with an approximate
coefficient $\overline H_{00}$, calculated with the spherical
weight function $w(\Omega)=1/4\pi$ and a 3-point integration.
We note that the exact procedure includes only 9 more 
terms with $l=2$.

\subsection{Treatment of the anisotropy in reciprocal-space approaches}
\label{sec:reciprocalCodes}

For completeness, we present here a simple recipe to take the 
anisotropy into account in reciprocal-space approaches.
The analytic function that appears in the computation 
of, e.g., the self-energy matrix element
$\langle\phi_{n\vq}|\Sigma(\omega')|\phi_{n\vq}\rangle$ 
is given by an expression of the form
\begin{equation}
a_{\bf GG'}(\vk) = \int \mathrm d^3 q \sum_{m}
[M^{mn}_{\bf G}(\vq,\vk)]^* M^{mn}_{\bf G'}(\vq,\vk) 
F(\omega,\omega',\epsilon_{m\vq-\vk})
\; ,
\end{equation}
where $M^{mn}_{\bf G}(\vq,\vk)=\langle\phi_{n\vq-\vk}|e^{-\mathrm i(\vk+\vG)\vr}|\phi_{n\vq}\rangle$
and $F(\omega,\omega',\epsilon_{m(\vq-\vk)})$
contains prefactors and frequency integrals \cite{Lebegue}.
For body and wings the approach outlined in the previous sections also applies
for this analytic function.
For the head element, on the other hand, the integral to compute is
\begin{equation}
\int\limits_{\mathrm Z_\Gamma} \!\!\mathrm d^3 k~ 
\frac{a_{\bf 00}(\vk)}{\kLk{}}
\; ,
\end{equation}
where $a_{\bf 00}(\vk)$ is analytic at $\vk = \mathbf 0$.
Assuming that $a_{\bf 00}(\vk)$ is non-zero and varies 
sufficiently slowly with $\vk$, we
set $a_{\bf 00}(\vk) \approx a_{\bf 00}(\mathbf 0)$ and 
evaluate the remaining integral in polar coordinates
\begin{equation}
\int\limits_{\mathrm Z_\Gamma} \!\!\mathrm d^3 k~ 
\frac{a_{\bf 00}(\mathbf 0)}{\kLk{}}
= a_{\bf 00}(\mathbf 0) \int \! \mathrm d\Omega_\vk~\frac{1}{\kLk{\hat}}
\int\limits_0^{k_{\rm max}(\Omega_\vk)} \!\!\!k^2 \mathrm dk~\frac{1}{k^2}
\; .
\end{equation}
In accordance with the computation of $\overline H_{00}$ as described in 
Appendix \ref{app:angularExpansion}, 
the angular integral can be computed numerically on an appropriate 
angular grid with the angular weight function
\begin{equation}
K(\Omega_\vk) = \int\limits_0^{k_{\rm max}(\Omega_\vk)}\!\!\! k^2 \mathrm dk~ \frac{1}{k^2}
= k_{\rm max}(\Omega_\vk)
\; .
\end{equation}
Alternatively, but formally equivalent, $K(\Omega_\vk)$ 
can be expanded in spherical harmonics with coefficients $K_{lm}$.
The integral then becomes
\begin{equation}
\int\limits_{\mathrm Z_\Gamma} \!\!\mathrm d^3 k~ 
\frac{a_{\bf 00}(\mathbf 0)}{\kLk{}}
= a_{\bf 00}(\mathbf 0) \sumlm K_{lm}^*  H_{lm}
\label{eq:Klm_hlm}
\; .
\end{equation}

Equation (\ref{eq:Klm_hlm}) is conducive for a discussion of different 
an\-iso\-tropy and the Coulomb singularity treatments.
In all isotropic approximations the sum over $l$ and $m$ is restricted
to the $l=0,m=0$ term. In the ``spherical'' approximation 
\cite{PhillipsKleinman} used in the early days of modern $GW$ calculations 
\cite{HybertsenLouie,GodbySchlueterSham} $K_{00}$ is further replaced 
by a shape-independent term
\begin{equation}
K^{\rm sph}_{00} = \frac{1}{Y_{00}} 
\left(\int \! \mathrm d\Omega ~k^3_{\rm max}(\Omega) \right)^{1/3}
\;. 
\end{equation}
The numerical computation of $H_{00}$ is restricted to 
3 angular points if the average over the Cartesian directions is taken,
which might introduce an additional inaccuracy for anisotropic systems.
In the improved integration scheme used by Pulci et al. \cite{Pulci} as
well as in the integration scheme by Wenzien et al. 
\cite{Wenzien}, the head of the inverse dielectric matrix is
written as a tensor and hence includes also the $l=2$ terms
(cf. Appendix \ref{app:angularExpansion}).
The tensor itself is chosen to reproduce the correct value in
the main directions, so $H_{lm}$ is effectively determined from
three independent points only.
In the scheme that we propose here only the choice of the angular grid determines
the accuracy of the sum in Equation (\ref{eq:Klm_hlm}).
It should be comparable to the scheme proposed by Hott \cite{Hott},
which formally includes all terms and also involves a numerical 
integration, the details of which are unfortunately 
not specified in \cite{Hott}.
Similarly, the offset-$\Gamma$-point method described in \cite{Lebegue}
in principle allows to capture the anisotropy to arbitrary precision.
However, the accuracy with which this is achieved in practice 
depends on the choice of $\vk$-points.

\section{Results}
\label{sec:results}
The equations presented in the previous sections were implemented 
into the \texttt{gwst} code \cite{Rieger,gwstGL}.
The test system was chosen to be a periodically repeated
4-layer Si(001) slab saturated with hydrogen and a vacuum separation
between the Si surface atoms equivalent to 4 layers. 
15 points per half-axis were used both
for the imaginary time and frequency Gauss--Legendre grids 
at a maximum numerical range of 6 atomic units. 
Convergence in the plane-wave cutoff is achieved for 7 Hartree, and 
unoccupied states up to 5 Hartree above the Fermi energy were included 
(610 bands).
Head and wings were computed separately with a $\vk$-point grid
of $14\times14\times1$ and 120 bands.
These parameters are sufficient to obtain quasiparticle energies 
converged to within 0.05\,eV.

Taking the body corrections from Equation (\ref{eq:body_result})
into account changes the quasiparticle
energies of our test system only little compared to 
neglecting them completely and setting
$\tilde\varepsilon_{\bf GG'}^{-1}({\bf 0}) = B^{-1}_{\bf GG'}$.
The magnitude of the corrections depends on the weight of the
$\Gamma$-point and hence on the $\vk$-point sampling, amounting
to $\sim$10\,meV for a $3\times3\times1$ sampling and
1--2\,meV for $8\times8\times1$. While these corrections are
small compared to the accuracy of our test calculation, they
might be larger for systems with stronger local-field effects.
Since the additional computational effort is small we always include them.

The repeated-slab arrangement considered here is a hypothetical system, 
that, at least until now, cannot be prepared in experiment.
However, since it is fundamentally a strongly anisotropic system, it provides 
an ideal test case for our modifications. It also provides the
possibility to tune the degree of anisotropy by varying the slab thickness and
the separation. If the slab separation were increased to infinity, the limit
of an isolated slab would be recovered. If additionally the slab thickness were
taken to infinity the limit of a hydrogenated silicon surface would be 
reached.
For the present choice of slab thickness and separation,
the non-zero elements of the dielectric 
tensor, including local-field effects and the contributions of
the non-local pseudopotential (cf. Appendix \ref{app:nonlocal_momentum}),
are $\varepsilon_{xx}$=5.1, $\varepsilon_{yy}$=5.5, and $\varepsilon_{zz}$=2.2
at the smallest imaginary frequency $\omega$=0.036 Hartree,
in agreement with effective-medium theory.
Without the contributions of the non-local pseudopotential
the values are $\varepsilon_{xx}$=5.9, $\varepsilon_{yy}$=6.5, 
and $\varepsilon_{zz}$=2.8,
which underlines their importance for our test system.
Varying the thickness or the separation produces changes in the 
quasiparticle energies that are of similar magnitude as the errors
from an inadequate treatment of the anisotropy. In order to be able to 
investigate the surface or isolated-film limit, it hence proved to be 
essential to take the anisotropy modifications into account \cite{PhilippPhD}.

\subsection{Convergence with respect to $l_{\rm max}$}
\label{sec:conv_L}

\begin{table}
\caption{Dependence of the quasiparticle energies (in eV 
relative to the valence-band maximum) on the
maximum angular momentum $l$ in Equation~(\ref{eq:Wl_in_r}).
The $\vk$-point sampling
for the data presented here is $4\times4\times4$.
Other samplings show the same behaviour.}
\label{tab:lmax}
\begin{tabular}{l|d{3}d{3}d{3}d{3}}
$l_{\rm max}$ & 0 & 2 & 4 & 6 \\\hline
lowest valence state    & -10.448 & -10.338 & -10.336 & -10.336 \\
lowest conduction state & 3.560 & 3.424 & 3.409 & 3.410 \\
\end{tabular}
\end{table}

In Table~\ref{tab:lmax} we report the convergence of the quasiparticle
energies with respect to the maximum value of $l$ used for the 
evaluation of $\Wlr$ in real space according to 
Equation~(\ref{eq:Wl_in_r}). It can be seen that already with 
$l_{\rm max}=2$ the results lie within our level of accuracy 
($\sim$ 0.05\,eV), and with $l_{\rm max}=4$ absolute convergence 
is reached.
These results also indicate that previous approaches \cite{Pulci,Wenzien},
which have treated the anisotropy at the level of $l_{\rm max}=2$,
have incorporated the most important aspects of the anisotropy 
since the terms from $l>2$ yield only minor corrections.
Nevertheless, as the computational cost for evaluating higher 
terms in Equation~(\ref{eq:Wl_in_r}) is negligible, we
use $l_{\rm max}=6$ in practice. 

\subsection{Convergence behaviour with respect to $\vk$-points}
\label{sec:conv}
\begin{figure}
\begin{center}
(a)
\includegraphics[scale=0.35,angle=-90,clip,trim=100 10 10 20]{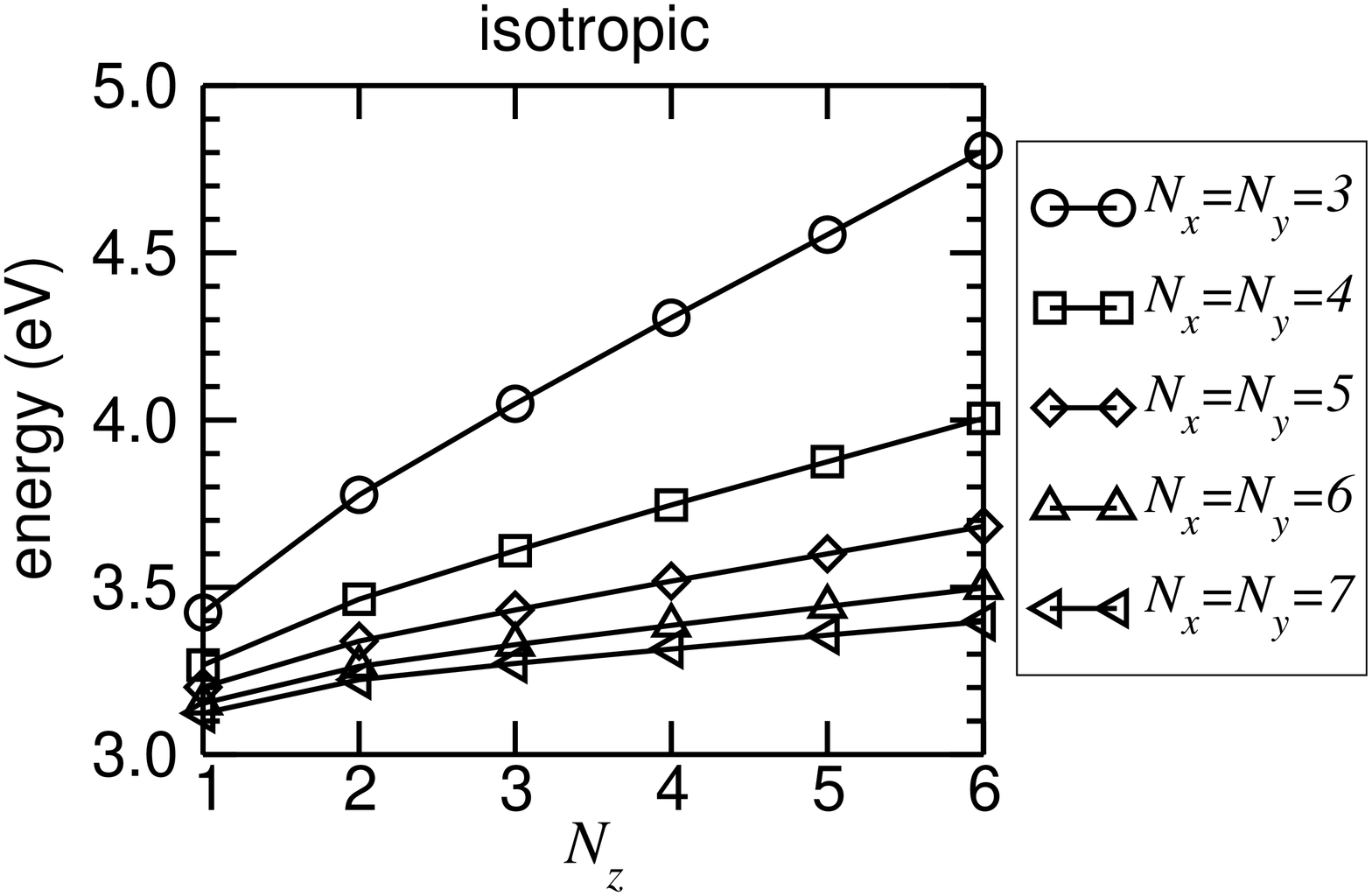}

(b)
\includegraphics[scale=0.35,angle=-90,clip,trim=100 10 10 20]{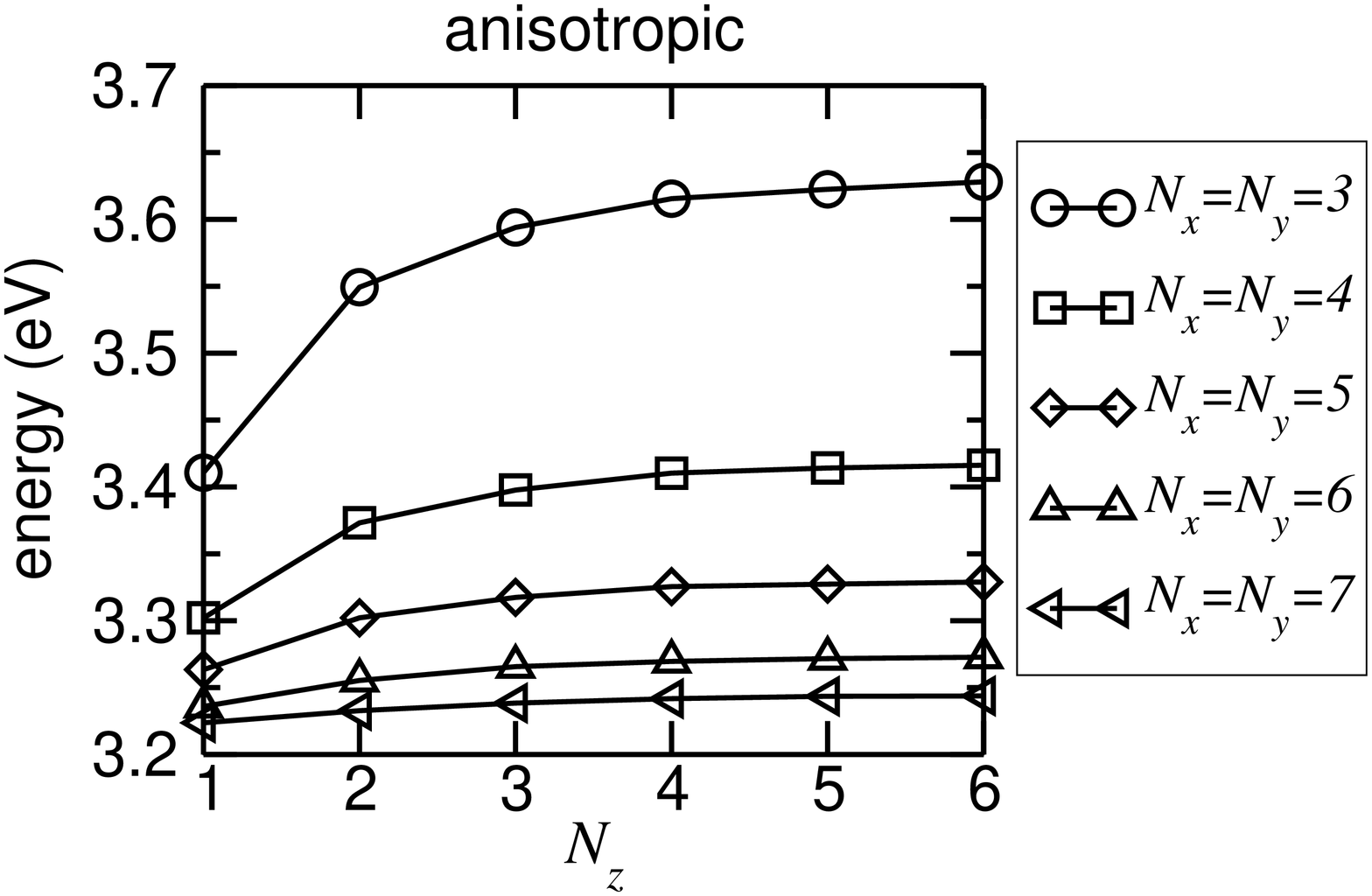}
\end{center}
\caption{Convergence of the lowest conduction-band energy 
with respect
to the number of k-points $N_z$ perpendicular to the surface for 
(a) the original isotropic implementation and (b) with the 
anisotropy taken into account.
The quantitative behaviour depends on the sampling in the 
parallel direction ($N_x=N_y$).
Note the different scales of the two graphs.
}
\label{fig:kz-conv}
\end{figure}

In Figure~\ref{fig:kz-conv} we show the convergence with respect to
the number of $\vk$-points in the direction perpendicular to 
the surface for the quasiparticle energy of the lowest conduction state. Other
states exhibit a similar behaviour. 
It is obvious that the original isotropic averaging for the 
screened interaction, notably in the long-range part, leads to an 
unphysical linear increase in the quasiparticle energy. In contrast,
the anisotropic treatment converges rapidly.

\begin{figure}
\includegraphics[scale=0.5,angle=-90,trim=50 40 70 40,clip]{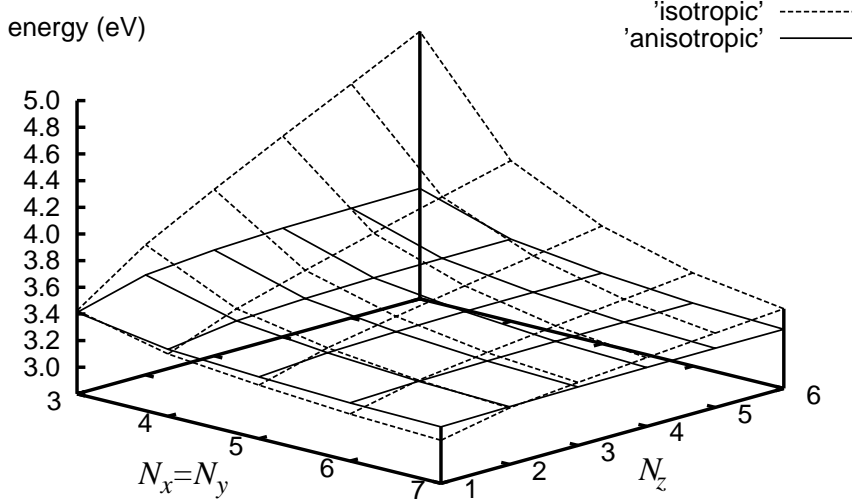}
\caption{Convergence of the lowest conduction-band state in the
isotropic and anisotropic treatment for various $\vk$-meshes.}
\label{fig:gnuplot}
\end{figure}

The reason for the linear increase in the isotropic treatment is found
in the inadequate treatment of the singularity, 
which is not fully removed. Integrating $1/|\bf k|^2$ 
numerically yields for $k_x\!=\!k_y\!=\!0$ with 
$\Delta k_z = k_{\rm max}/N_z$
\begin{equation}
\sum_{n=1}^{N_z} \Delta k_z \frac{1}{(n  \Delta k_z)^2}
~\longrightarrow~
\int\limits_{\Delta k_z}^{k_{\rm max}}\!\! \mathrm dk~ \frac{1}{k^2} 
= \frac{1}{\Delta k_z} - \frac{1}{k_{\rm max}}
= \frac{N_z-1}{k_{\rm max}}
\end{equation}
and hence a linearly diverging contribution, whose weight
is proportional to $\Delta k_x\Delta k_y \sim (N_xN_y)^{-1}$. 
When the $\vk$-sampling is increased in all three directions simultaneously,
no such linear divergence occurs, as can be seen from the
three-dimensional plot in Figure~\ref{fig:gnuplot} when
going from the left side (small $N_x,N_y,N_z$) to the right. 
However, such a restriction is undesirable and inefficient in practice.
Therefore, only the proper anisotropic treatment enables us to investigate
the importance of the $\vk$-point sampling in the direction perpendicular to 
the surface, which is directly related to the interaction with adjacent slabs
in $GW$ calculations \cite{PhilippPhD}.
To our knowledge, the convergence in the perpendicular direction has
not been addressed in previous $GW$ calculations for slab systems, 
probably under the erroneous assumption that neighbouring slabs 
do not interact.

The computed quasiparticle energies appear to be a linear function of
the product $\Delta k_x \Delta k_y$ when $N_z$ is kept fixed (not shown).
This can be exploited to extrapolate the value for 
$\Delta k_x=\Delta k_y \rightarrow 0$, as shown in
table~\ref{tab:extrap}.
\begin{table}
\caption{Extrapolated $\Delta k_x=\Delta k_y \rightarrow 0$ 
quasiparticle energies in eV for the
lowest conduction state for the isotropic and anisotropic treatment
with a fixed number of $\vk$-points in the $z$-direction $N_z$.}
\label{tab:extrap}
\begin{tabular}{l|d{3}d{3}d{3}d{3}d{3}d{3}}
$N_z$ & 1 & 2 & 3 & 4 & 5 & 6 \\
\hline
isotropic & 3.063 & 3.094 & 3.090 & 3.082 & 3.074 & 3.064 \\
anisotropic & 3.179 & 3.160 & 3.158 & 3.157 & 3.157 & 3.156 
\end{tabular}
\end{table}
The extrapolated values evidently depend much less on the number
of $\vk$-points in the $z$-direction than those for finite $\Delta k_{x}$ and
$\Delta k_y$.
For isotropic averaging, the situation
greatly improves after extrapolation, but a small, systematic trend 
towards lower energies for larger $N_z$ is still present.
This indicates that even after extrapolation no reliable convergence with respect 
to  the number of $\vk$-points can be reached in the isotropic case.
A comparison with the anisotropic treatment shows that the absolute
error of the isotropic averaging is about 0.1\,eV, larger than could 
have been estimated from the isotropic data alone.

\section{Summary}
\label{sec:conclusions}

We have presented a comprehensive account of the treatment of anisotropic screening
in $GW$ calculations that employ reciprocal space for the computation of
the screened interaction. 
In particular, we have demonstrated that this requires only small modifications 
of the original $GW$ space-time implementation \cite{Rieger}. 
The additional terms are computationally not very demanding. Furthermore, we
have shown that the treatment of the anisotropy in other $GW$ implementations can be
understood in terms of approximations to the exact equations. 

The improvements presented in this article greatly increase the efficiency of
$GW$ calculations for anisotropic systems in the space-time method, e.g., for
films and surfaces.
This is mostly due to the fact that the fully anisotropic treatment enables us
to converge the $\vk$-point sampling in the perpendicular and parallel direction
separately, whereas isotropic averaging leads to an unacceptable
linear divergence in this case. The number of $\vk$-points required for
converged results is thus reduced considerably.


\begin{ack}
We thank Lucia Reining, Fabien Bruneval, Francesco Sottile, Carlo Rozzi, 
Hardy Gross, Angel Rubio, and Christoph Friedrich
for fruitful discussions. This work was funded in part by the EU
through the Nanophase Research Training Network (Contract No. 
HPRN-CT-2000-00167) and the Nanoquanta Network of Excellence 
(Contract No. NMP-4-CT-2004-500198).
Philipp Eggert acknowledges the Deutscher Akademischer Austauschdienst
for financial support.
\end{ack}

\appendix
\section{Angular expansion of vector expressions}
\label{app:angularExpansion}
In this section we briefly summarise how simple
expressions for a normalised vector $\hat \vk$ can be written in terms 
of spherical harmonics of the corresponding spatial angle $\Omega_\vk$.
The expansion of a scalar product $\hat \vk\cdot\vr$ 
requires spherical harmonics of order $l=1$
\begin{equation}
  \hat\vk\cdot\vr = \sum_{m=-1}^1 r_m Y_{1m}(\Omega_\vk)
\end{equation}
with
\begin{equation}
  r_0 = \sqrt{\frac{4\pi}{3}} r_z \;, \quad
   r_{\pm1} = \sqrt{\frac{2\pi}{3}} (\mp r_x + \mathrm i r_y)
\; .
\end{equation}

It is also straightforward to show that a tensor expression
$\kLk{\hat}$, where $\bf L$ is symmetric, can be 
written in terms of spherical harmonics up to $l$=2 as
\begin{equation}
\kLk{\hat} = \sum_{l\in \{0,2\}} \sum_{m=-l}^{l}
L_{lm} Y_{lm}(\Omega_\vk)
\end{equation}
with the coefficients

\begin{equation}
\begin{array}{rclcrcl}
 \displaystyle 
  L_{00}   & = & \displaystyle \sqrt{\frac{4\pi}{9}}      (L_{xx} + L_{yy} + L_{zz})  
  \;,
 & \quad & 
 \displaystyle
  L_{20} & = & \displaystyle \sqrt{\frac{4\pi}{45}}   (2 L_{zz} - L_{xx} - L_{yy}) 
\; ,\\
  L_{2,\pm 1} & = & \displaystyle \sqrt{\frac{8\pi}{15}}  (\mp L_{xz} + \mathrm i L_{yz}) 
\; ,
 & \quad &
 \displaystyle
  L_{2,\pm 2} & = & \displaystyle \sqrt{\frac{2\pi}{15}}   (L_{xx} - L_{yy} \mp 2\mathrm i L_{xy})
\; .
\end{array}
\end{equation}

In the $GW$ space-time method the term $\kLk{\hat}$ appears in the denominator
of the head element of the inverse dielectric 
matrix (\ref{eq:head_aniso})  as well as its product with
an angular weight function $w(\Omega_\vk)$. In order to expand 
these expressions in spherical harmonics 
\begin{eqnarray}
\frac{1}{\kLk{\hat}}&=& 
\sumlm H_{lm} Y_{lm}(\Omega_\vk)
\; ,
\\
\frac{w(\Omega_\vk) }{\kLk{\hat}}&=& 
\sumlm \overline H_{lm} Y_{lm}(\Omega_\vk)
\; ,
\end{eqnarray}
we determine the coefficients numerically by performing the following
integrals on a Lebedev--Laikov grid \cite{LebedevLaikov}
\begin{eqnarray}
H_{lm} &=& \int\! \mathrm d\Omega_\vk~ Y^*_{lm} (\Omega_\vk)
 \frac{1}{\kLk{\hat}}
\; ,
\\
\overline H_{lm} &=& \int\! \mathrm d\Omega_\vk~ Y^*_{lm} (\Omega_\vk)
\frac{w(\Omega_\vk)}{\kLk{\hat}}
\label{eq:Hlm_bar}
\; .
\end{eqnarray}
Since $\kLk{}$ and $w(\Omega_\vk)$ are
even functions, only even $l$-components contribute to the sums.

\section{Dielectric matrix}
\label{ap:dielec}
For calculating the long-range limit of the symmetrised dielectric matrix
we follow the derivation of Baroni and Resta \cite{BaroniResta}.
For $\vk\rightarrow {\bf 0}$ we have 

\begin{eqnarray}
P_{\bf 00}(\vk) &\sim  |k|^2 &\qquad \textnormal{``head''}
\label{eq:Phead},\\
P_{\bf G0}(\vk) &\sim  |k|^{~} &\qquad \textnormal{``wing''},
\\
P_{\bf GG'}(\vk) & &\qquad \textnormal{``body''}
\label{eq:Pbody}
\;.
\end{eqnarray}
This behaviour holds even for the exact polarisability of a non-metallic 
system and cancels the Coulomb singularity in the symmetrised 
dielectric matrix. 
In the context of the $GW$ approximation 
Equations (\ref{eq:Phead}) to (\ref{eq:Pbody}) are also valid for 
full self-consistency. However,
   here we restrict ourselves 
to the non-self-consistent case and derive expressions for
the corresponding Taylor coefficients from
the Adler--Wiser formula \cite{Adler,Wiser}
for the polarisability
\begin{eqnarray}
P_{\bf GG'}(\vk, {\rm i}\omega) &=& - \frac{4}{(2\pi)^3}\sum_{v,c}
\int \!\mathrm d^3 q ~
\frac{\epsilon_{c\vq+\vk} - \epsilon_{v\vq}}
{(\epsilon_{c\vq+\vk} - \epsilon_{v\vq})^2 + \omega^2}
\\&&\times
\nonumber
\left<\varphi_{v\vq}\Big|e^{-{\rm i}(\vk+\vG)\cdot\vr}\Big|\varphi_{c\vq\!+\!\vk}\right>
\left<\varphi_{c\vq\!+\!\vk}\Big|e^{{\rm i}(\vk+\vG')\cdot\vr}\Big|\varphi_{v\vq}\right>
\; ,
\end{eqnarray}
where the sum over $v$ and $c$ runs over occupied and unoccupied states, respectively.
For $\vG'={\bf 0}$ and $\vk \rightarrow {\bf 0}$ this leads to
\begin{eqnarray}
P_{\bf G0}(\vk, {\rm i}\omega) &\rightarrow& 
- \frac{4{\rm i}}{(2\pi)^3}\sum_{v,c} \int \!\mathrm d^3 q ~
\frac{\epsilon_{c \vq} - \epsilon_{v \vq}}
{(\epsilon_{c \vq} - \epsilon_{v \vq})^2 + \omega^2}
\\&&\times
\nonumber
\left<\varphi_{v\vq}\Big|e^{-\mathrm i\vG\cdot\vr}\Big|\varphi_{c\vq}\right>
\left(\vk \cdot 
\left<\varphi_{c\vq}|\vr|\varphi_{v\vq}\right>
\right)
\; ,
\end{eqnarray}
while for $\vG=\vG'={\bf 0}$ and $\vk \rightarrow {\bf 0}$ the result is
\begin{eqnarray}
P_{\bf 00}(\vk, {\rm i}\omega) &\rightarrow& - \frac{4}{(2\pi)^3}\sum_{v,c}
\int \!\mathrm d^3 q ~
\frac{\epsilon_{c \vq} - \epsilon_{v \vq}}
{(\epsilon_{c \vq} - \epsilon_{v \vq})^2 + \omega^2}
\\&&\times
\nonumber
\left(\vk \cdot \left<\varphi_{v\vq}|\vr|\varphi_{c\vq}\right>\right)
\left(\vk \cdot \left<\varphi_{c\vq}|\vr|\varphi_{v\vq}\right>\right)
\; .
\end{eqnarray}
The computation of the matrix elements $\left<*|\vr|*\right>$ for
the Kohn--Sham eigenfunctions is presented in
Appendix~\ref{app:nonlocal_momentum}.
Writing the scalar products \mbox{$\vk \cdot \left<*|\vr|*\right>$}
as \mbox{$\sum_\alpha k_\alpha \left<*|r_\alpha|*\right>$}, where
$\alpha$ runs over the spatial (Cartesian) directions, we arrive at
the following expressions for the wings of the symmetrised dielectric matrix
in terms of a new vector quantity $\mathbf U(\vG,\omega)$:
\begin{eqnarray}
\tilde\varepsilon_{\bf G0}(\vk, {\rm i}\omega) &\rightarrow& 
\sum_\alpha \frac{k_\alpha}{|\vk|}  
\frac{16\pi i}{(2\pi)^3 |\vG|}\sum_{v,c}
\int \!\mathrm d^3 q ~
\frac{\epsilon_{c \vq} - \epsilon_{v \vq}}
{(\epsilon_{c \vq} - \epsilon_{v \vq})^2 + \omega^2}
\\&&\nonumber \times
\left<\varphi_{v\vq}\Big|e^{-\mathrm i\vG\cdot\vr}\Big|\varphi_{c\vq}\right>
\left<\varphi_{c\vq}|r_\alpha|\varphi_{v\vq}\right>
\\&=:&
\sum_\alpha \frac{k_\alpha}{|\vk|}  U_\alpha(\vG,\omega)
\label{eq:eps_wing}
\end{eqnarray}
and analogously for the head element
in terms of a new tensor quantity $\mathrm F(\omega)$
\begin{eqnarray}
\tilde\varepsilon_{\bf 00}(\vk, {\rm i}\omega) &\rightarrow& 1 + 
\sum_{\alpha,\beta} \frac{k_\alpha k_\beta}{|\vk|^2}
\frac{16\pi}{(2\pi)^3}\sum_{v,c}
\int \!\mathrm d^3 q~ 
\frac{\epsilon_{c \vq} - \epsilon_{v \vq}}
{(\epsilon_{c \vq} - \epsilon_{v \vq})^2 + \omega^2}
\\&&\times
\nonumber
\left<\varphi_{v\vq}|r_\alpha|\varphi_{c\vq}\right>
\left<\varphi_{c\vq}|r_\beta|\varphi_{v\vq}\right>
\\
&=:&
\sum_{\alpha,\beta} \frac{k_\alpha k_\beta}{|\vk|^2}
 F_{\alpha\beta}(\omega)
\label{eq:eps_head}
\; .
\end{eqnarray}
Since for many systems head and wings converge much slower with respect to the
$\vk$-point sampling, but faster with respect to the number of conduction bands
compared to the body \cite{BaroniResta}, we compute them in a separate
calculation.
Equations (\ref{eq:eps_wing}) 
and (\ref{eq:eps_head}) hold also for the self-consistent case,
but the coefficients $U_{\alpha}(\vG; \omega)$ and 
$F_{\alpha\beta}(\omega)$ would have to be computed differently.
The considerations following from now on then apply
to self-consistent $GW$, too.

Equations~(\ref{eq:eps_wing}) and~(\ref{eq:eps_head}) illustrate that
the limit for $\vk\rightarrow {\bf 0}$ is finite but will, in general, depend
on the direction $\vk/|\vk|=\hat \vk$ in which the 
$\Gamma$-point is approached.
We denote this directional dependence in the limit $\vk ={\bf 0}$
by the directional (spatial) angle $\Omega_\vk$.
In existing implementations
the treatment of the directional dependence varies.
Sometimes, the direction $\Omega_\vk$ is simply fixed
to a single value.
It is also common to carry the directional dependence through the
inversion by performing a block-wise inversion \cite{PickCohenMartin},
which is also the approach taken in the original space-time implementation
\cite{Rieger}.
For brevity, we omit the frequency argument from the following derivation.
We denote the body of the symmetrised dielectric matrix 
($\vG\ne0, \vG'\ne 0$) at $\vk ={\bf 0}$ by $\mathbf B$, the wings
by $\mathbf w$ (a column vector) and $\mathbf w^\dagger$ (a row vector), 
and the head by $H$. The symmetrised dielectric matrix hence takes the
form
\begin{equation}
\tilde\varepsilon(\Omega_\vk) = 
\left(
\begin{array}{cc}
H(\Omega_\vk) & \mathbf w^\dagger(\Omega_\vk)\\
\mathbf w(\Omega_\vk) & \mathbf B \\
\end{array}\right)
\; .
\label{eq:defineHWB}
\end{equation}
Head, wings, and body
of the symmetrised inverse dielectric matrix are then given by
\cite{PickCohenMartin}
\begin{eqnarray}
\tilde \varepsilon^{-1}_{\bf 00}(\Omega_\vk) & = & 
\Big[ H(\Omega_\vk) - 
\sum_{\bf G,G'\ne0}
       w_{\vG}^*(\Omega_\vk)  
       B^{-1}_{\bf GG'}  w_{\vG'}(\Omega_\vk)
\Big]^{-1}
\label{eq:epsi_head}
\; ,\\
\tilde \varepsilon^{-1}_{\bf G0}(\Omega_\vk) & = & 
- \tilde \varepsilon^{-1}_{\bf 00}(\Omega_\vk)  
       \sum_{\vG'\ne\mathbf 0} 
       B_{\bf GG'}^{-1} w_{\vG'}(\Omega_\vk)
\; ,\\
\tilde \varepsilon^{-1}_{\bf GG'}(\Omega_\vk) & = & 
B^{-1}_{\vG \vG'}
+ \tilde \varepsilon^{-1}_{\bf 00}(\Omega_\vk)  
\Big[ 
       \sum_{\vG''\ne\mathbf 0} 
       B^{-1}_{\bf GG''} w_{\vG''}(\Omega_\vk)
\Big]
\\&&\qquad\times
\nonumber
\Big[ 
       \sum_{\vG''\ne \mathbf 0} 
        w_{\vG''}^*(\Omega_\vk) B^{-1}_{\bf G''G'}
\Big]
\; .
\end{eqnarray}
Using equation~(\ref{eq:eps_wing}), we now define the auxiliary vector
\begin{equation}
S_\alpha(\vG) = \sum_{\vG'\ne \mathbf 0} 
B^{-1}_{\vG\vG'}
U_\alpha(\vG') 
\label{eq:definitionS}
\end{equation}
and rewrite equation~(\ref{eq:epsi_head}) as
\begin{equation}
\tilde \varepsilon^{-1}_{\bf 00}(\Omega_\vk)  =  
\Big[ \sum_{\alpha,\beta} \hat k_\alpha \hat k_\beta  \Big(
      F_{\alpha\beta} - 
       \sum_{\bf G\ne0} U^*_\alpha(\vG) S_\beta(\vG)
       \Big)
\Big]^{-1}
=: \frac{1}{\kLk{\hat}}
\; ,
\end{equation}
thus defining $\mathbf L$. Correspondingly, we have
\begin{eqnarray}
\tilde \varepsilon^{-1}_{\bf G0}(\Omega_\vk) & = & 
- \frac{ \hat \vk \cdot \mathbf{S}(\vG) }{\kLk{\hat }}
\; ,\\
\tilde \varepsilon^{-1}_{\bf GG'}(\Omega_\vk) & = & 
B^{-1}_{\vG \vG'}
+  \frac{ \left[\hat \vk \cdot \mathbf{S}(\vG)\right]
          \left[\hat \vk \cdot \mathbf{S}^*(\vG')\right]}
        {\kLk{\hat}}
\; .
\end{eqnarray}

\section{Kleinman--Bylander correction to the matrix elements of the
position operator}
\label{app:nonlocal_momentum}

The matrix elements of $\vr$, which enter the expressions in the 
previous section, are in practice calculated via
the commutator of $\bf r$ with the Kohn--Sham Hamiltonian $h^{\rm KS}$ as
\begin{equation}
\langle \varphi_{c\vq} | \vr | \varphi_{v\vq} \rangle
=
\frac{\langle \varphi_{c\vq} | [h^{\rm KS}, \vr] | \varphi_{v\vq} \rangle}
{\epsilon_{c\vq} - \epsilon_{v\vq}}
\; .
\end{equation}
While the contribution from the kinetic-energy operator is
trivial to compute, that from the non-local
pseudopotential $V_{\rm nl}$ is more cumbersome and has
often been neglected in earlier calculations. We
show here that it is possible to compute it
efficiently in a separable expression.

In its separable Kleinman--Bylander form \cite{KleinmanBylander} 
the non-local pseudopotential operator is written in the Dirac notation as
\begin{equation}
V_{\rm nl} = \sum_\mu |\phi_\mu\rangle
                       \frac{1}{E_{\mu}} \langle\phi_\mu|
\;,
\end{equation}
where $\mu$ is a collective index $\{\mathbf R_\mu, n_\mu, l_\mu, m_\mu\}$ 
that runs over all pseudopotential projectors 
while $\phi_\mu$ is in general given in a radial basis around a certain atomic
position $\mathbf R_\mu$, i.e.,
\begin{equation}
\phi_\mu(\mathbf r)= f_{n_\mu l_\mu}(|\mathbf r-\mathbf R_\mu|) 
 Y_{\!l_\mu m_\mu}(\Omega_{\mathbf r-\mathbf R_\mu})
\; .
\end{equation}
In addition $\mu$ can run over chemical species, which does not alter 
the following derivation, except that $f_{nl}$ then also depends on the 
species.
We will now show that the commutator can be factorised,
which reduces the scaling to be linear in the number of plane waves instead of
quadratic as demonstrated in a previous approach \cite{VnlOld}.
To this end we consider the commutator of $\vr$ with a single projector
\begin{eqnarray}
&&
 \Big(|\phi_\mu\rangle\frac{1}{E_\mu}\langle\phi_\mu| {\mathbf r}\Big)
~-~
\Big({\mathbf r} |\phi_\mu\rangle\frac{1}{E_\mu}\langle\phi_\mu|\Big)
\nonumber\\ 
&=& 
   \frac{1}{E_\mu}
\Big[  
    |\phi_\mu\rangle\langle\phi_\mu|
       ({\mathbf r}-\mathbf R_\mu)
~-~  ({\mathbf r}-\mathbf R_\mu)  
    |\phi_\mu\rangle\langle\phi_\mu|
  \Big]
  \; .
\end{eqnarray}
Next we make use of the fact that ${\mathbf r} - \mathbf R_\mu$ can 
be expressed in the same radial basis as $\phi_\mu$
\begin{equation}
[{\mathbf r} - \mathbf R_\mu]_\alpha = |\mathbf r- \mathbf R_\mu| 
 \sum_{m=-1}^1 c_{\alpha m} Y_{\!1m} (\Omega_{\mathbf r- \mathbf R_\mu})
\; ,
\end{equation}
where $\alpha \in \{ x, y, z \}$ are the spatial directions, and 
$c_{\alpha m}$ yield the spatial components of the spherical harmonics
for $l=1$:
\begin{center}
\begin{tabular}{l|ccc}
$c_{\alpha m}$ & $\alpha=x$ & $\alpha=y$ & $\alpha=z$ \\
\hline
$m=-1$ & $\frac 1{\sqrt2}$ & $\frac i{\sqrt2}$ & 0 \\
$m=0$  & 0 & 0 & 1 \\
$m=1$ & $-\frac 1{\sqrt2}$ & $\frac i{\sqrt2}$ & 0
\end{tabular}
\end{center}
We can then write the product in the radial basis, too,
\begin{eqnarray}
    |\phi_\mu^\alpha\rangle &:=&
[{\mathbf r} - \mathbf R_\mu]_\alpha |\phi_\mu\rangle\nonumber\\
& = & |\mathbf r- \mathbf R_\mu| \sum_{m=-1}^1 c_{\alpha m} 
      Y_{\!1m} (\Omega_{\mathbf r- \mathbf R_\mu})
f_{n_\mu l_\mu}(|\mathbf r-\mathbf R_\mu|) 
 Y_{\!l_\mu m_\mu}(\Omega_{\mathbf r-\mathbf R_\mu})
   \nonumber\\
   &=& 
\sum_{L=|l_\mu-1|}^{l_\mu+1}\sum_{M=-L}^{L} c_{\alpha M, m_\mu}^{L,l_\mu}  
f^{\mathrm r}_{n_\mu l_\mu}(|\mathbf r\!-\! \mathbf R_\mu|)
Y_{\!L M}(\Omega_{\mathbf r- \mathbf R_\mu})
\end{eqnarray}
with
\begin{eqnarray}
   f^{\rm r}_{n l}(\rho) &=& \rho  f_{n l}(\rho) \; ,
   \\
   c_{\alpha M, m}^{L,l} &=& \sum_{m'} c_{\alpha m'} 
 (lm~ 1m'|L M)
 \; ,
\end{eqnarray}
where $(l m~ 1 m'|L M)$ is a Clebsch--Gordan coefficient.
It is convenient to express $\phi^\alpha_\mu$ in a
plane-wave basis similar to what is done for $\phi_\mu$. If the 
functions $f_{nl}$ are given on a radial grid \cite{fhi98PP},
$f^{\rm r}_{nl}$ is trivial to compute, and the same routines that are used 
to compute $\phi_\mu(\Gk)$ in the DFT calculation
can be employed for
the summands in $\phi^\alpha_\mu(\Gk)$.
It must be emphasised that the sums over $L$ and $M$ contain only a 
very small number of non-zero terms (at most six).

The final formula is thus again a separable expression
\begin{equation}
\left[V_{\rm nl}, r_\alpha\right] = \sum_\mu 
    \frac{1}{E_\mu} \left(
      \big|\phi_\mu\big>\langle\phi^\alpha_\mu|
    - |\phi^\alpha_\mu\rangle\big<\phi_\mu\big|
    \right)
    \label{eq:nl_result}
\; .
\end{equation}
The computational effort to set up a full
$N_v \times N_c$ matrix for all three directions requires
$4N_\vG N_\mu (N_v + N_c)$ operations to calculate the
$\langle\phi_\mu|\varphi_{v/c}\rangle$ and
$\langle\phi^{\alpha}_\mu|\varphi_{v/c}\rangle$ projections
and $6 N_\mu N_v N_c$ operations to build up the 3 matrices 
from the projections in Equation~(\ref{eq:nl_result}). The scaling is thus linear in the
number of $\vG$-vectors $N_\vG$ and not quadratic \cite{VnlOld}.

We have tested the size of the contributions from the non-local part
of the pseudopotential for GaN and the II-VI compounds ZnO, ZnS, and CdS. 
We found that the macroscopic dielectric constant changes between +8 and --15\%, 
 which results in changes of the quasiparticle energies between --0.03\,eV and 0.15\,eV 
\cite{Rinke/etal:2005}.
\bibliographystyle{./elsart-num-notitle}
\bibliography{./aniso-gwst}

\end{document}